\documentclass{aa}
\usepackage{txfonts}
\usepackage{graphicx}
\usepackage{natbib}

\bibpunct{(}{)}{;}{a}{}{,}

\usepackage{color}

\newcommand{\kms}{\,{\rm km\,s}^{-1}}

\newcommand{\Lsun}{L_\odot}

\newcommand{\Msun}{\,\mathrm{M}_\odot}

\newcommand{\as}{\ifmmode {^{\scriptscriptstyle\prime\prime}}
        \else $^{\scriptscriptstyle\prime\prime}$\fi}

\newcommand{\simless}{\mathbin{\lower 3pt\hbox
      {$\rlap{\raise 5pt\hbox{$\char'074$}}\mathchar"7218$}}} 
\newcommand{\simgreat}{\mathbin{\lower 3pt\hbox
     {$\rlap{\raise 5pt\hbox{$\char'076$}}\mathchar"7218$}}} 

\newcommand{\tabletwo}{
\begin{table}
\caption{Dust disk modeling results.}
\begin{tabular}{llll}
\hline
Parameter & Value & Unit & \\
\hline
$p$        &  $-0.49 \pm 0.02$  &    & Surface density exponent \\
$H_0$      &  $12.7 \pm 0.3$  & au & Scale height at 100 au  \\
$h$        &  $-0.34 \pm 0.04$  &    & exponent of scale height \\
$R_\mathrm{out}$ &  $187.0 \pm0.1$ & au & Outer radius \\
$i$        & $90.8 \pm 0.4$ & $^\circ$ & Inclination \\
$PA$       & $2.98 \pm 0.02$ & $^\circ$ & Orientation \\
\hline
\end{tabular}
  \label{tab:disk}
  \tablefoot{$\Sigma(r)=\Sigma_0(r/100\,\mathrm{au})^{-p}$,
  $H(r) = H_0 (r/100\,\mathrm{au})^{-h}$, and $T(r) = T_0 (r/100\,\mathrm{au})^{-q}$
  with $q=0.4$. Errors are 1 $\sigma$.}
\end{table}
}

\begin{document}

\title{The shadow of the Flying Saucer:\\
A very low temperature for large dust grains.
}
\author{
S. Guilloteau \inst{1,2}, V. Pi\'etu \inst{3}, E. Chapillon \inst{1,2,3}, E. Di Folco \inst{1,2}, A. Dutrey\inst{1,2},
T.Henning\inst{4}, D.Semenov\inst{4}, T.Birnstiel \inst{4} and N.Grosso\inst{5}
}
\institute{
Univ. Bordeaux, LAB, UMR 5804, F-33270 Floirac, France
\and
CNRS, LAB, UMR 5804, F-33270 Floirac, France\\
  \email{[guilloteau,difolco,dutrey]@obs.u-bordeaux1.fr}
\and
IRAM, 300 rue de la piscine, F-38406
Saint Martin d'H\`eres, France
\and
Max Planck Institute f\"ur Astronomie, K\"onigstuhl 17, D-69117 Heidelberg, Germany
\and
Observatoire Astronomique de Strasbourg, Universit\'e de Strasbourg, CNRS, UMR 7550, 11 rue de l'Universit\'e, 67000 Strasbourg, France
}

\offprints{S.Guilloteau, \email{Stephane.Guilloteau@obs.u-bordeaux1.fr}}

\date{Received 23 October 2015 / Accepted 21 December 2015} %
\authorrunning{Guilloteau et al.} %
\titlerunning{The shadow of the Flying Saucer}

\abstract
{Dust determines the temperature structure of protoplanetary disks, however, dust temperature determinations almost invariably rely on
a complex modeling of the Spectral Energy Distribution.
}
{We attempt  a direct determination of the temperature of large grains
emitting at mm wavelengths.}
{We observe the edge-on dust disk of the Flying Saucer,
which appears in silhouette against the CO J=2-1 emission from a background
molecular cloud in $\rho$ Oph. The combination of velocity gradients due
to the Keplerian rotation of the disk and intensity variations in the CO
background as a function of velocity allows us to directly measure the 
dust temperature. The dust opacity can then be derived from
the emitted continuum radiation.}
{The dust disk absorbs the radiation from the CO clouds at several
velocities. We derive very low dust temperatures, 5 to 7 K at radii around 100 au,
which is much lower than most model predictions. The dust optical depth is $> 0.2$ at 230 GHz,
and the scale height at 100 au
is at least 8 au (best fit 13 au). However, the dust disk is very flat (flaring index -0.35),
which is indicative of dust settling in the outer parts.}
{}


\maketitle{}

\section{Introduction} Although dust is the main agent to control the
protoplanetary disk temperature \citep{Chiang+Goldreich_1997}, our
knowledge of dust temperatures essentially relies on modeling of disk
images and SED \citep[e.g.,][]{Dalessio+etal_2001}. Despite (or even
because of) their sophistication, these models suffer from many
uncertainties because of the large number of assumed properties: radial
distribution, dust grain growth, dust settling, composition and
porosity, disk flaring geometry, etc. The model temperatures are
mostly constrained by the near-IR part of the SED, which comes from the
optically thick region within 10 au from the star, and dust
temperatures in the disk midplane only rely on the strong extrapolation
that the model provides radially and vertically. Gas temperatures can
provide a proxy for the dust temperature, but because molecules become
heavily depleted on grains at (dust) temperatures below 15 K, this
method only samples the lukewarm region 1 to 2 scale heights above the
disk plane.

We present  a completely different method for a direct measurement of the dust
temperature. The method relies on seeing a disk in silhouette against a bright background.
Edge-on disks are the best targets for this because their larger line-of-sight
opacity maximizes the expected signal.

The \object{Flying Saucer} (\object{2MASS J16281370-2431391})
is an isolated, edge-on disk in the outskirts
of the $\rho$ Oph clouds \citep{Grosso+etal_2003} with
evidence for large dust grains \citep{Pontoppidan+etal_2007}.
\citet{Grosso+etal_2003} resolved  the light scattered by (micron) dust grains in near-infrared with the NTT and
the VLT and estimated a
disk radius of $2.15''$, which is about 260 au for the adopted distance of 120 pc
\citep{Loinard+etal_2008}. The detection of the CN N=2-1 line \citep{Reboussin+etal_2015}
confirms the existence of a large gas disk. The $\rho$ Oph region
is crowded with molecular clouds that are brightly emitting in CO lines. However, the
low extinction derived by \citet{Grosso+etal_2003} toward the Flying Saucer
suggests it lies in front of these clouds, providing an ideal geometry
for our purpose. We thus observed the Flying Saucer at high angular
resolution in CO J=2-1 with the Atacama Large Millimeter Array (ALMA)
and the molecular clouds with the IRAM 30-m telescope.

\section{Observations}
\paragraph{IRAM 30-m}
We obtained a CO J=2-1 spectrum of the Flying Saucer with the IRAM
30-m to measure the brightness of the (foreground or background) emission.
Pointing was performed on Saturn, which was a few degrees above the Flying
Saucer at the time of the observations.
The weather was relatively windy, which combined with the low declination
of the Flying Saucer, resulted in some anomalous refraction on angular
scales of a few arcsec.

\begin{figure}
   \centering
   \includegraphics[width=0.85\columnwidth]{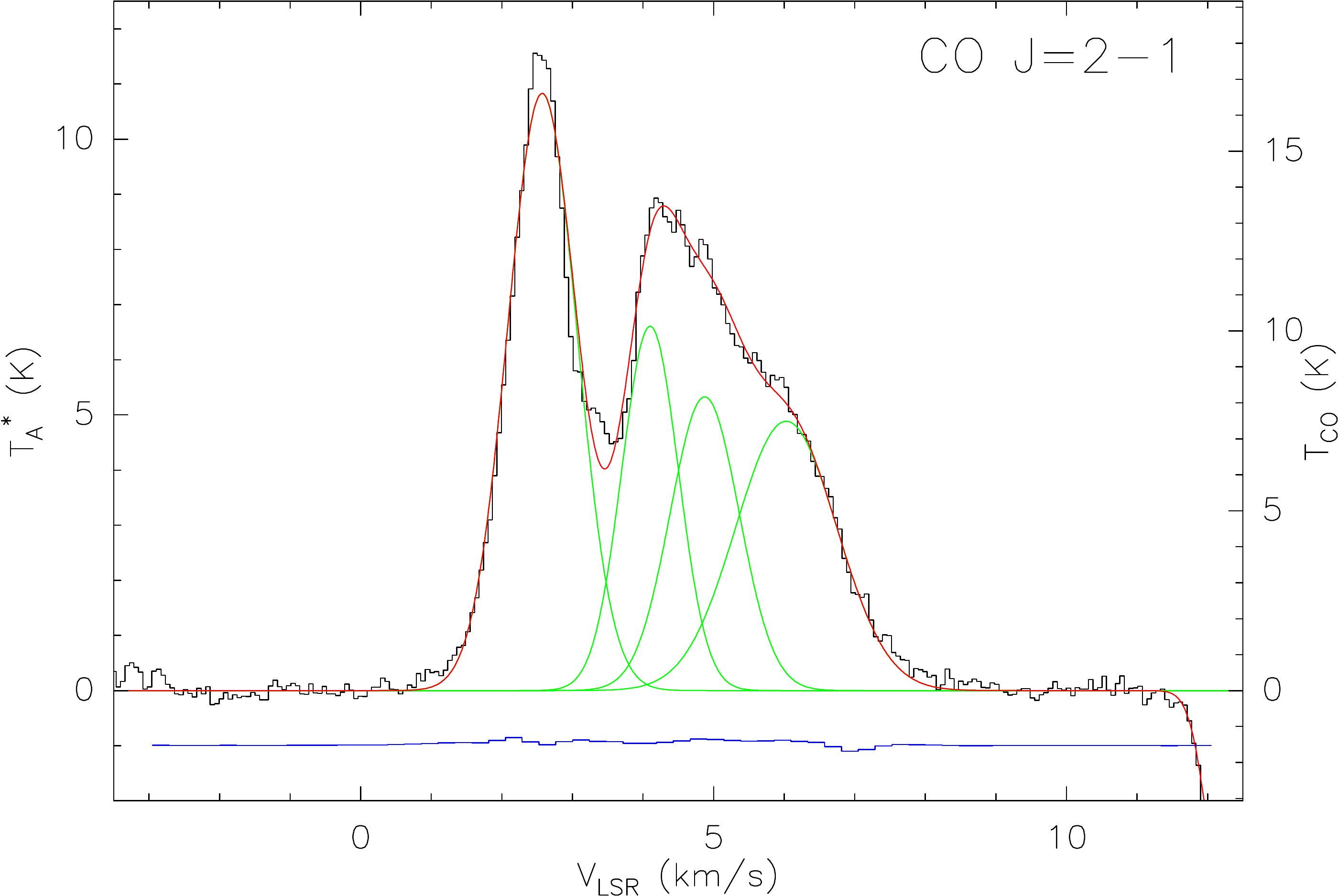}
 \caption{CO J=2-1 spectrum toward the Flying Saucer in a $11''$ beam (black).
 A decomposition in four Gaussians is overlaid. The blue line is the
 integrated disk spectrum derived from the ALMA observations (shifted for
 clarity). The negative feature at $12 \kms$ is tropospheric CO.
 The right axis scale is the brightness temperature.}
  \label{fig:spectrum}
\end{figure}

At 230 GHz, the beamsize is $10.7''$ and the beam efficiency
is $0.60,$ while the forward efficiency is $0.92$. The conversion
from $T_A^*$ to Jy is thus 7.6 Jy/K.
The pointing issues mentioned above  broaden
the effective beamsize, but should not significantly
affect the conversion from the antenna temperature scale
to sky brightness temperature because we are interested in
sources extended compared to the 30-m beam size.
We first performed an on-off measurement, using a reference position
at $(1998'',846'')$, in a region of low emission selected with the
FCRAO CO J=1-0 wide field survey (\footnote{COMPLETE team, 2011,
FCRAO Ophiuchus 12CO cubes and map, http://hdl.handle.net/10904/10078,
Harvard Dataverse, V2}). The reference position was observed in
frequency switching and the spectrum was added back to the on-off result,
providing excellent agreement with a frequency switch spectrum on
the Flying Saucer. 
The final 30-m spectrum (Fig.\ref{fig:spectrum})
can be decomposed in four Gaussian profiles of typical line width around
$1 \kms.$

\begin{figure*}
    \centering
     \includegraphics[width=17.0cm]{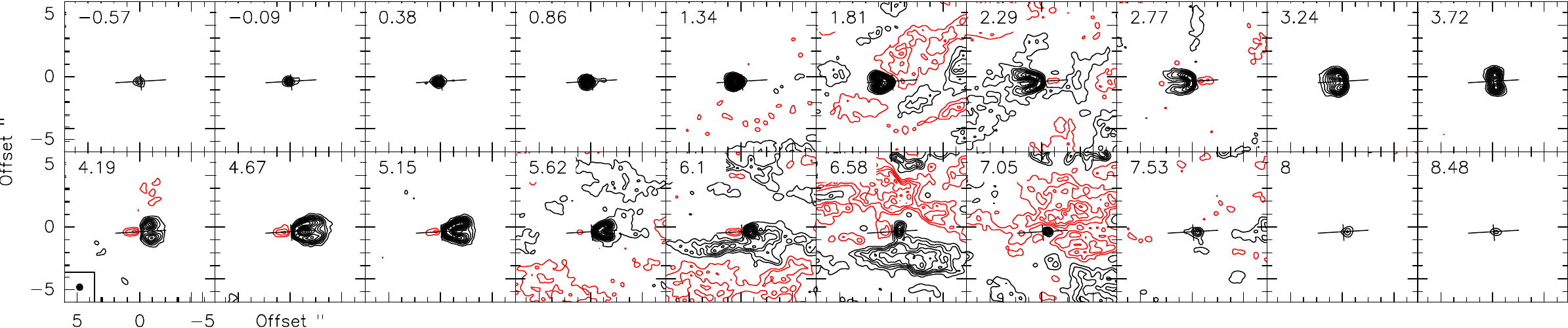}
    \caption{Channel maps of the $^{12}$CO J=2-1 line emission toward the Flying Saucer.
    Contours are in step of 8 mJy/beam (0.76 K); negative contours are red. Velocities
    are indicated in each panel. The cross indicates the position and orientation of
    the dust disk.}
    \label{fig:co-map}
\end{figure*}

\paragraph{ALMA}
The high angular resolution observations of CO J=2-1 were performed
with ALMA on 17 and 18 May 2015 under excellent weather (Cycle 2, 2013.1.00387.S).
The correlator was configured to
deliver very high spectral resolution with a channel spacing of
15 kHz (and an effective velocity resolution of 40 m/s).
Data was calibrated via the standard ALMA calibration script in the CASA software
package. Titan was used as a flux calibrator. The calibrated data
was regridded in velocity to the LSR frame and exported
through UVFITS format to the GILDAS package for imaging and data analysis.
No self-calibration was performed. With robust weighting, the $uv$ coverage
provided by the $\sim 34$ antennas yields a circular beamsize of 0.5$''$.

Figure \ref{fig:co-map} present channel maps of the CO J=2-1 emission.
An enlarged version of the central channels is presented in Fig.\ref{fig:co-fine}.
In Fig.\ref{fig:montage}, we show the dust continuum emission, the
integrated CO map, the integrated CO J=2-1 spectrum, and a
position-velocity diagram through the disk plane. The total continuum
flux is 35 mJy at 242 GHz (with 5\% calibration uncertainty).

\section{Results and analysis}
The dust disk clearly extends out to about 180 au, while the gas disks spreads
to 300 au. The CO data is highly contaminated by emission from molecular clouds,
which results in a complex integrated spectrum over the whole disk
(Fig.\ref{fig:montage}d). At velocities around $\sim 1.8 \kms$ and in the
range $6-7 \kms$, the filamentary structure of these clouds  appears
clearly(see Fig.\ref{fig:co-map}).

\begin{figure}
    \centering
     \includegraphics[width=0.9\columnwidth]{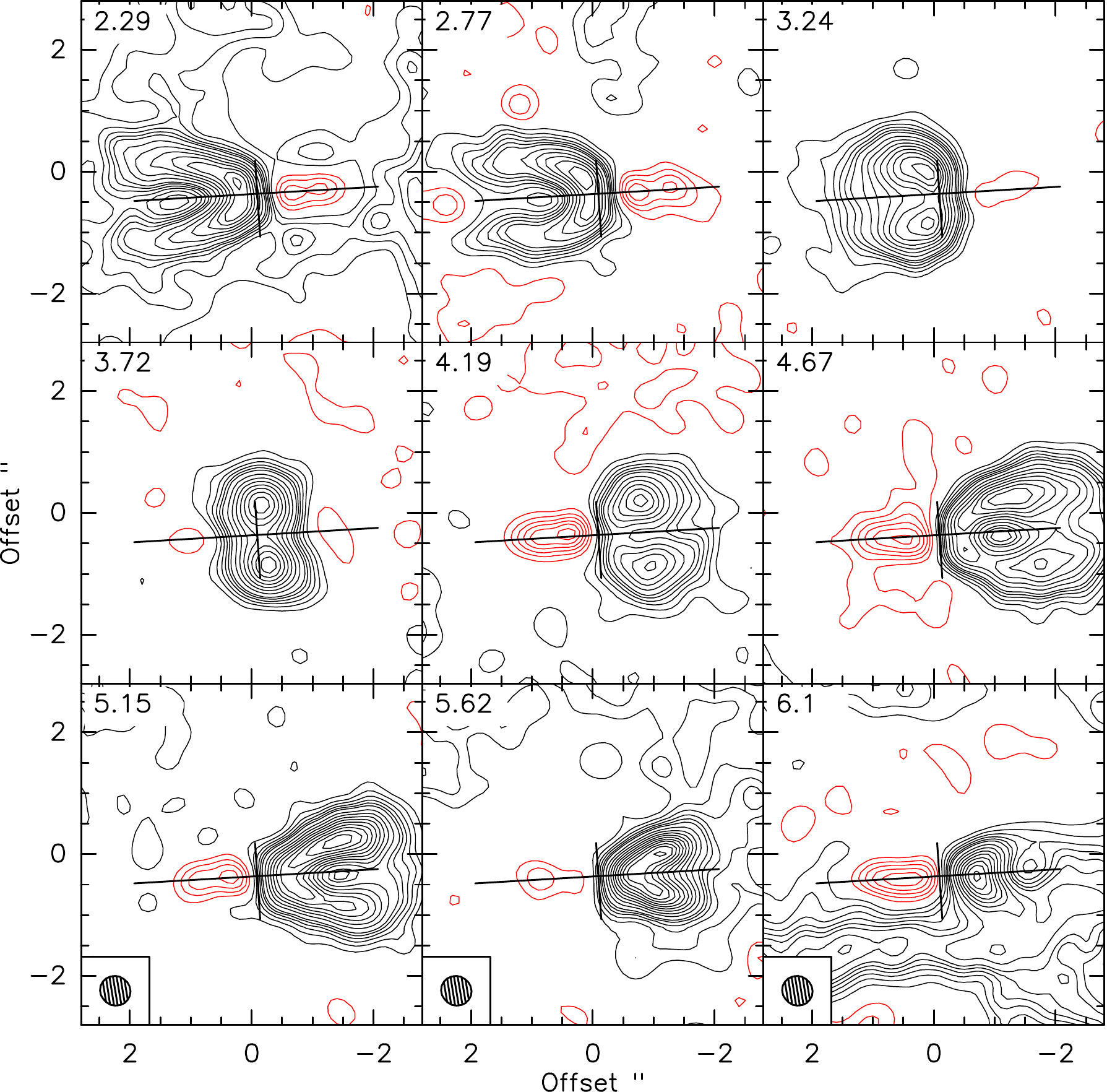}
    \caption{Channel maps of the $^{12}$CO J=2-1 line emission toward the Flying Saucer.
    Contours are in steps of 4 mJy/beam (0.38 K, approximately 2.7 $\sigma$) up
    to 28 mJy/beam, and 8 mJy/beam above; negative contours are red. The
    apparent absorption peaks at six contour levels in the east, and five in the West.}
    \label{fig:co-fine}
\end{figure}

\begin{figure}
    \centering
     \includegraphics[width=0.9\columnwidth]{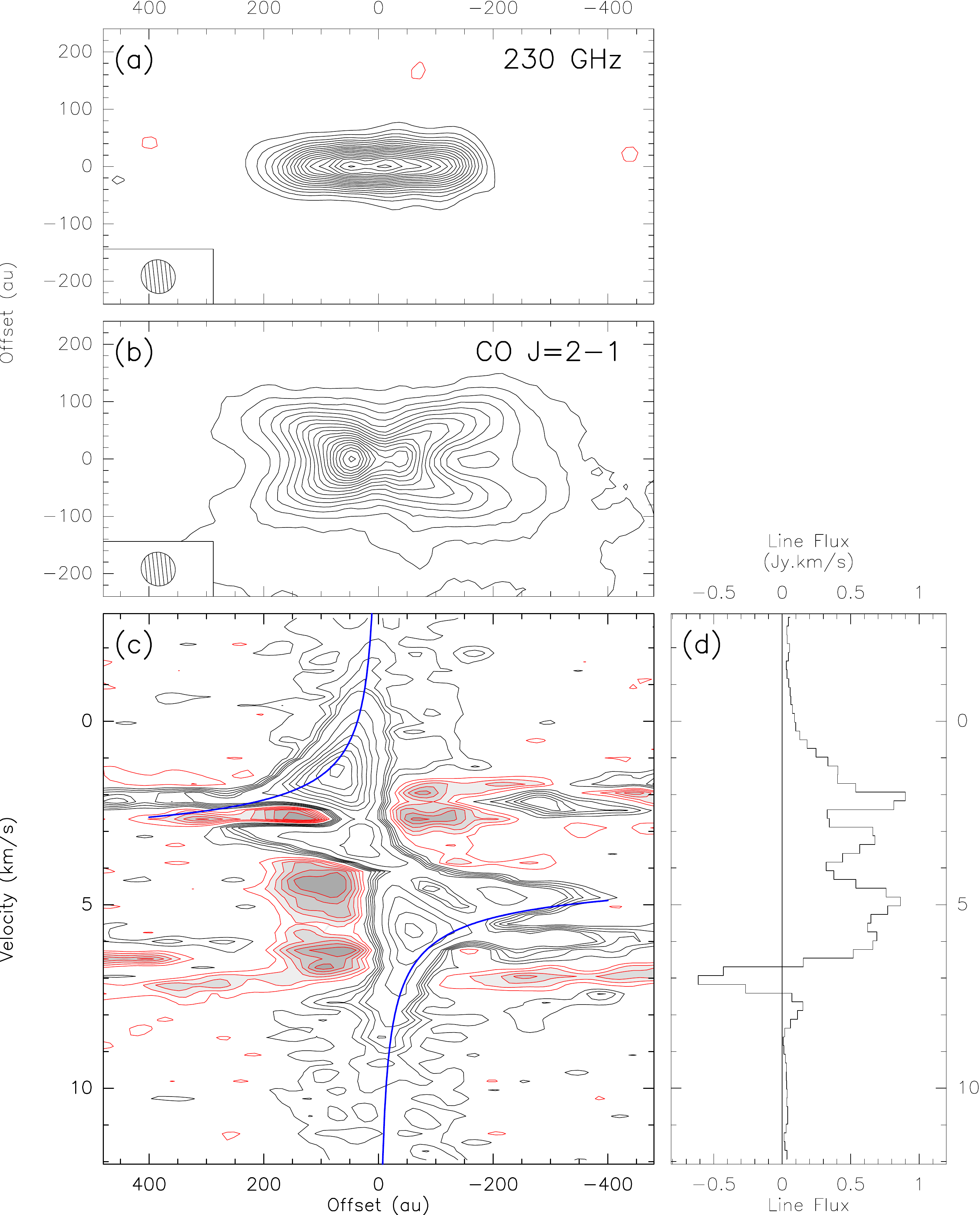}
    \caption{
    (a) Continuum emission at 230 GHz; contour spacing is 0.4 mJy/beam.
    (b) Integrated CO line flux; contour spacing is 20 mJy/beam\,km/s.
(c) Position-velocity diagram across the disk plane; contour spacing
    as in Fig. \ref{fig:co-fine}. The blue curve is the Keplerian
    velocity for a $0.57 \Msun$ star.
    (d) Integrated CO spectrum over the disk.
    Images have been rotated by $3^\circ$ clockwise to align the disk along the x-axis.}
    \label{fig:montage}
\end{figure}

Around an LSR velocity of $5 \kms$, CO in the western part shows the pattern
expected from Keplerian rotation,
and clearly reveals that CO originates from the upper layers of the disk
contrary to the mm-emitting dust that has settled toward the midplane.
In the eastern part, however, CO shows absorption
against apparently nothing (see Fig.\ref{fig:co-fine}).
This is a result of interferometric filtering of extended emission from
these molecular clouds. The results depend on whether
the cloud is in the foreground \citep[see][their Fig.7]{Gueth+etal_1997} or background.
In the first case, the dust disk can be hidden by optically thick CO
from the cloud, leading to null visibilities if that cloud is
sufficiently spatially extended ($>7.6''$ with our shortest baselines
of 21 m).

However, in Figs.\ref{fig:co-map}-\ref{fig:co-fine},
the signal goes to negative values, requiring
a different explanation.
A smooth extended CO emission is filtered out by the interferometer whatever
its opacity, leading to a null brightness.
A continuum dust
source located in front of the CO cloud absorbs this CO emission
if the dust temperature is below that of the background CO cloud.
The resulting brightness at the CO velocity toward the dust disk is then  smaller than around it,
reaching negative values since the
background extended emission is filtered out by the interferometer.

At frequencies away from that of the CO line, the apparent dust brightness in a synthesized beam is
\begin{equation}
T_d = f (1-\exp(-\tau)) (J_\nu(T_\mathrm{dust}) - J_\nu(T_{bg}))
,\end{equation}
where $\tau$ is the opacity; $f$ the beam filling factor; and
$J_\nu$, the radiation temperature, is the Planck function multiplied by $c^2/2k\nu^2$,
\begin{equation}
J_\nu(T) = \frac{h \nu}{k} \frac{1}{\exp(h\nu/(kT))-1} 
,\end{equation}
and $T_{bg}$ is the cosmological background, $J_\nu(T_{bg}) = 0.19$ K at 230 GHz.
At velocity $\mathrm{v}$, the CO cloud has a brightness
\begin{equation}
T_c(\mathrm{v}) =  (1-\exp(-\tau_c(\mathrm{v}))) (J_\nu(T_\mathrm{cloud}) - J_\nu(T_{bg}))
.\end{equation}
Thus, at velocity $\mathrm{v}$, the brightness of the dust disk becomes
\begin{equation}
T_l(\mathrm{v}) = f (1-\exp(-\tau)) (J_\nu(T_\mathrm{dust}) - T_c(\mathrm{v}))
.\end{equation}
Subtracting  $T_d$ and $T_l$ yields the opacity and beam filling factor product
\begin{equation}
 f (1-\exp(-\tau)) = \frac{T_d - T_l(\mathrm{v})}{ T_c(\mathrm{v}) - J_\nu(T_\mathrm{bg})}
\label{eq:fill}
\end{equation}
from which we derive
\begin{equation}
J_\nu(T_\mathrm{dust}) =  \frac{T_d (T_c(\mathrm{v}) - J_\nu(T_\mathrm{bg})) }{ T_d - T_l(\mathrm{v}) } + J_\nu(T_{bg})
\label{eq:tdust}
,\end{equation}
which, as $J_\nu(T_\mathrm{bg}$) is small, is about
\begin{equation}
J_\nu(T_\mathrm{dust}) \approx  \frac{T_d}{ T_d - T_l(\mathrm{v}) } T_c(\mathrm{v})
\label{eq:tprop}
.\end{equation}
From Fig. \ref{fig:montage}, $T_d$ is typically 0.4\,K,  while $T_l(\mathrm{v})$ is
$\sim -2.0$\,K around $4 \kms$ and $\sim -1.5$\,K at 2.8 and 5-6 $\kms$.
Using $T_c(\mathrm{v})$ from Fig.\ref{fig:spectrum}, 8 to 17 K, 
we thus derive (average) dust temperatures as low as 5 to 8 K.
From Eq. \ref{eq:fill}, the filling factor $f$ is at least 0.2, which for a beam size
of $0.5''$ (60 au) indicates a minimum FWHM of the dust disk of 13 au along the minor axis.
Given the FWHM of $\sim 200$\,au along the major axis, the aspect
ratio is consistent with an inclination $\sim 87^\circ$, or implies a
minimal $1/e$ scale height of 8 au if the disk is purely edge-on.
Also, the dust opacity at 230 GHz must be $> 0.2$ and is
most likely substantially higher.
\begin{figure}
    \centering
     \includegraphics[width=0.76\columnwidth]{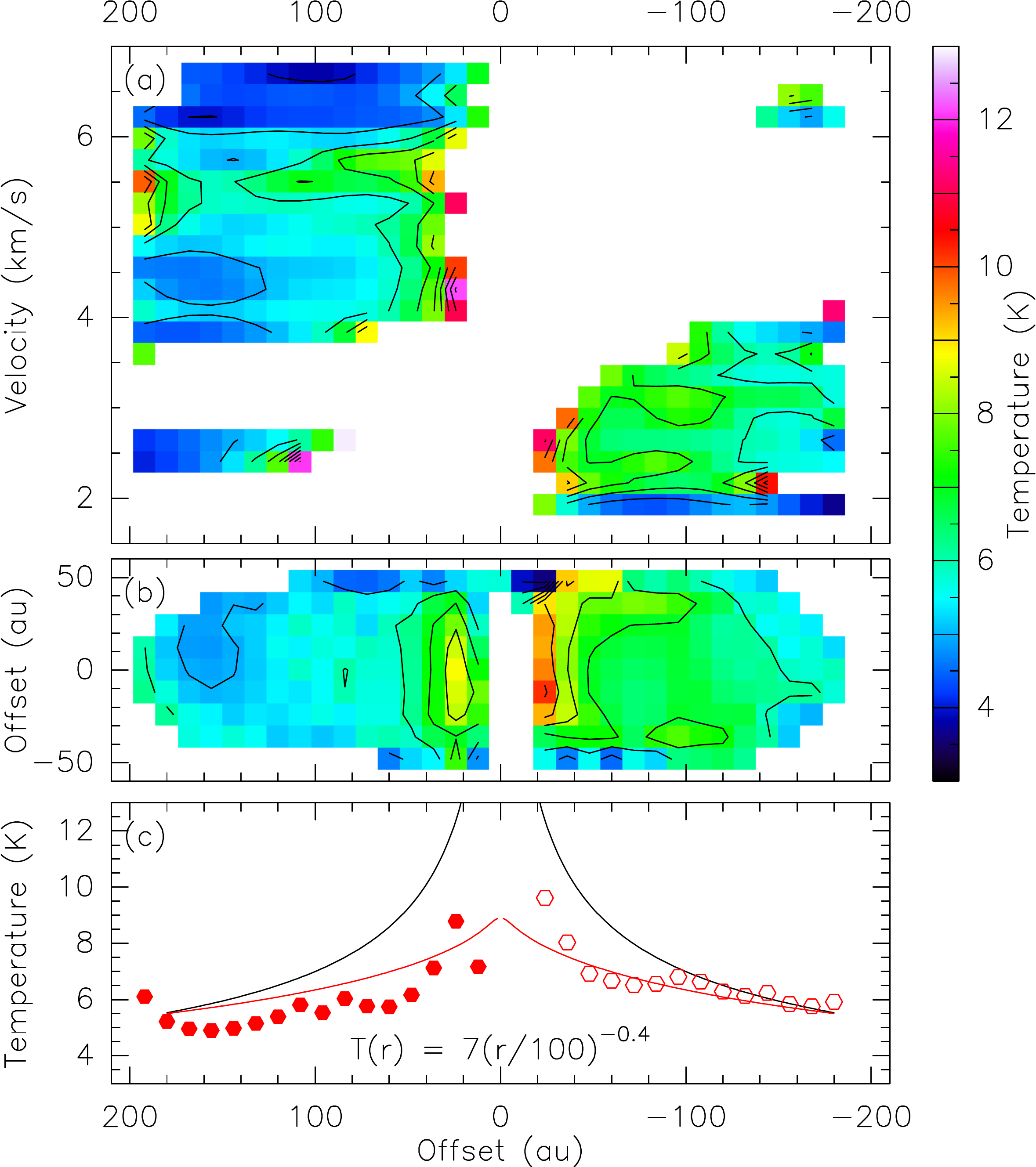}
    \caption{(a) Dust temperatures derived as a function of position (impact
    parameter in au) and velocity. (b) Map of the dust temperature, taken
    as the mean of all valid values derived for different velocities. The
    typical error on the mean is $\leq 1$ K. (c) Cut along the disk
    plane and comparison between the local temperature and
    the line-of-sight average (in red).}
    \label{fig:tdust}
\end{figure}
Eq. \ref{eq:tdust} can be applied for each pixel and velocity channel, allowing
us to derive a datacube of dust temperatures.
As the dust disk is unresolved vertically,
we show these temperatures
as a function of velocity along the disk axis in Fig. \ref{fig:tdust}a,
revealing the consistency of the derived values.
Residual structures in the CO clouds probably perturb the
measurement at LSR velocities around $5.6 \kms$and at the extreme
velocities where the clouds become optically thin and unveil their
spatial column density variations. Yet, the derived values are
remarkably consistent. Averaging the datacube over the whole velocity
range produces the dust temperature map of Fig. \ref{fig:tdust}b.
A small radial gradient is visible, but the inner regions ($< 40$\,au)
may be biased because of residual emission from the (warm) CO disk itself.
The derived temperature is an average along the line of sight, sampling radii
between the impact parameter and the disk radius of about 190 au.
This effect is estimated in Fig. \ref{fig:tdust}c, assuming
a uniform surface density in the disk. It shows the measured values are
consistent with $T(r)= 7\,\mathrm{K}\,(r/100\,\mathrm{au})^{-0.4}$.
This absorption derived temperature only relies
on the absolute calibration of the 30-m spectra (the ALMA
calibration factor cancels out in Eq. \ref{eq:tprop}), which we estimate
accurate to 10\%.

If the cloud has significant structure on scales 7-10$''$,
the local brightness $T_c(\mathrm{v})$ toward the Flying Saucer
can be higher (but also equally likely lower). The maximum expected
brightness can be estimated  assuming
the cloud is the maximum size not sampled in our interferometric maps,
7.6$''$, and computing the filling factor in the $10.7''$ beam of the 30-m
telescope.
This would increase $T_c(\mathrm{v})$ by a factor 1.9. However, because
of the nonlinearity of Eq. \ref{eq:tdust}, the dust temperature law
would then become $T(r)= 10\,\mathrm{K}\,(r/100\,\mathrm{au})^{-0.5}$.
Also,  this is a maximally biased correction since it assumes that
the all four different clouds (corresponding to the four Gaussian
lines in Fig. \ref{fig:spectrum}) have their brightness maximum
in the direction of the Flying Saucer.

\tabletwo

We analyzed the 242 GHz continuum with a simple disk model that is vertically
isothermal, using truncated
power laws for temperature and surface density and no dust settling
(tapered edge viscous profiles yielded a poorer fit).
We assumed $T = T_0 (r/100 \mathrm{au})^{-0.4}$, following \citet{Guilloteau+etal_2011}.
We ran an MCMC, using the ``emcee'' tool of \citet{Foreman-Mackey+etal_2013},
which implements the affine invariant sampler of \citet{Goodman+Weave_2010}
with 40 walkers and 5000 steps, retaining only the last 1000 steps for
the error analysis. Results are in Table \ref{tab:disk}.
The disk appears closer to edge-on than derived in the NIR analysis
of \citet{Grosso+etal_2003}.
The low flaring index and negative $p$ value are the direct
signatures of an edge-on settled dust disk fitted by a
disk model that does not consider dust settling, as demonstrated by \citet{Boehler+etal_2013}.
The two observing frequencies, 230 GHz and 242 GHz, are insufficiently
separated to provide constraints on the spectral index, and leave a
strong degeneracy between the dust temperature and surface density, so
no reliable value could be found. \citet{vanKempen+etal_2009}
report a flux of 58 mJy in the 15$''$ beam of SCUBA at 850$\mu$m, giving a spectral
index of 1.4 between 242 and 345 GHz. This would imply a mean dust
temperature of 10 K if the dust disk is optically thick,  7 K if
we take the minimum opacity at 230 GHz, 0.2, and a dust emissivity
exponent $\beta=0.8$. However, error bars are not specified and
the region has extended flux, so these estimates are not very reliable.
With standard disk dust properties \citep[from][with $\beta=0.8$]{Beckwith+etal_1990},
the disk H$_2$ surface density is at least
$10^{23}$ cm$^{-2}$ at 100 au.

All measurements thus point toward very low dust temperature.
More precise values can be obtained by measuring the missing short spacings
of the ALMA data, either via the Alma Compact Array or  on-the-fly
mapping at the 30-m, and by obtaining an accurate SED covering
the mm and sub-mm range, as the turnover frequency of the Planck function
should be readily visible.

\section{Discussion}
The very low temperatures are at odds with previous estimates.
Direct measurements of mm-emitting dust temperatures are rare.
\citet{Dutrey+etal_2014} report 14 K at 200 au in \object{GG Tau},
using deviations from the Rayleigh-Jeans part of the blackbody
between 230 and 690 GHz. However, the GG Tau disk is circumbinary and
its thermal structure is special because the puffed-up inner rim of
the tidal cavity effectively shades the outer ring.
This leads to a steep temperature gradient (almost as 1/r) that is not expected in simpler
circumstellar disks.
\citet{Guilloteau+etal_2011} argue that some
of the brighter disks have optically thick cores and used the apparent
disk brightness to derive the temperature profile (see their Table 8).
The temperatures of 15 to 25 K found for the cores of \object{DG Tau}, \object{MWC 480},
\object{T Tau}, \object{DG Tau B}, and \object{HL Tau} extrapolate to values around 11 to 19 K at
100 au when the mean exponent of 0.4 is considered. However, all these
stars are rather luminous and at least three of them are relatively massive (1.7 to 2 $\Msun$).

Most disk modeling also predicts rather high temperatures \citep[e.g.,][]{Dullemond+etal_2002}.
\citet{Andrews+etal_2013} argue for an average dust disk temperature of
$T_\mathrm{dust} = 25 (L_*/\Lsun)^{(1/4)} K$ for disks extending out to 200 au.
This is similar to the results found by \citet{Isella+etal_2009}. In particular,
for \object{DM Tau}, whose mass is very similar to that of the Flying Saucer,
\citet{Isella+etal_2009} derived a dust temperature of 20 K at 100 au

To yield lower temperatures than model predicts, either the disk must intercept much less
stellar light or the dust must be a more efficient emitter at longer wavelengths
than assumed.
The J,H,Ks images of the Flying Saucer from \citet{Grosso+etal_2003} show direct
evidence for a flared disk and require a stellar luminosity of $\sim 0.10 \Lsun$,
which is appropriate for a $0.57 \Msun$ star, and, hence, the later hypothesis is preferred.
\citet{Voshchinnikov+Semenov_2000} showed that for grains containing
a conducting material (refractory organics,
FeS, FeO, etc\ldots), nonspherical grains can be colder than the spherical grains by
~20-50\%. For stochastically grown aggregates made of silicates, \citet{Fogel+Leung_1998}
have found a similar (but weaker, 10-20\%) cooling effects.

Another possibility is a difference in temperature between larger grains, which dominate
at long wavelengths, and small grains, which dominate the SED in the NIR.
If the Planck mean opacity is lower than 1, larger grains are expected to be colder
than smaller grains. The difference in temperature can be reduced through thermal
accommodation by collisions with the gas, but the efficiency of the process is low
at densities below $10^{10}$cm$^{-3}$.

It is also worth pointing out that the dust emissivity may depend on the temperature
\citep[e.g.,][]{Boudet+etal_2005,Coupeaud+etal_2011}, an effect which may be
substantial given the very low values found in this study.
Finally, such very low dust temperatures, if applicable to all grain sizes,
 affect the disk chemistry by reducing the mobility of molecules on grains and
halting all surface reactions except hydrogenation by H atoms.

\begin{acknowledgements}
This work was supported by ``Programme National de Physique Stellaire'' (PNPS
from INSU/CNRS.)
This research made use of the SIMBAD database,
operated at the CDS, Strasbourg, France.
This paper makes use of the following ALMA data:
   ADS/JAO.ALMA\#2013.1.00387.S. ALMA is a partnership of ESO (representing
   its member states), NSF (USA), and NINS (Japan), together with NRC
   (Canada),  NSC and ASIAA (Taiwan), and KASI (Republic of Korea) in
   cooperation with the Republic of Chile. The Joint ALMA Observatory is
   operated by ESO, AUI/NRAO, and NAOJ.
This paper is based on observations carried out with the IRAM 30-m telescope.
IRAM is supported by INSU/CNRS (France), MPG (Germany), and IGN (Spain).
\end{acknowledgements}


\bibliography{biblio}
\bibliographystyle{aa}

\end{document}